\documentclass[conference,a4paper]{IEEEtran}

\usepackage{graphicx}              
\usepackage{amsmath}               
\usepackage{amsfonts}  
\usepackage{amssymb}              
\usepackage{amsthm}
\usepackage{epstopdf}
\usepackage{mathtools}
\usepackage{hyperref}

\newtheorem{theorem}{Theorem}[section]

\newcommand{\F}{\mathcal{F}} 
\newcommand{\I}{\mathbb{I}} 

\newcommand{\G}{\mathbf{G}} 
\newcommand{\R}{\mathbf{R}} 
\newcommand{\W}{\mathbf{W}} 

\newcommand{\A}{\mathbf{A}} 
\newcommand{\C}{\mathbf{C}} 

\newcommand{\B}{\mathbf{B}} 
\newcommand{\X}{\mathbf{X}} 
\newcommand{\Y}{\mathbf{Y}}

\newcommand{\spn}{\text{span}}
\DeclareMathOperator{\diag}{\text{diag}}
\newcommand*{\tm}[1]{{\scriptstyle[#1]}}
\begin{document}

\sloppy

\title{Achievability of Nonlinear Degrees of Freedom in Correlatively Changing Fading Channels}

\author{
  \IEEEauthorblockN{Mina Karzand}
  \IEEEauthorblockA{Massachusetts Institute of Technology\\
    Cambridge, USA\\
    Email: mkarzand@mit.edu} 
  \and
  \IEEEauthorblockN{Lizhong Zheng}
  \IEEEauthorblockA{Massachusetts Institute of Technology\\
    Cambridge, USA\\
    Email: lizhong@mit.edu}
}



\maketitle

\begin{abstract}

A new approach toward the noncoherent communications over the time varying fading channels is presented. In this approach, the relationship between the input signal space and the output signal space of a correlatively changing fading channel is shown to be a nonlinear mapping between manifolds of different dimensions. Studying this mapping, it is shown that using nonlinear decoding algorithms for single input-multiple output (SIMO) and multiple input multiple output (MIMO) systems, extra numbers of degrees of freedom (DOF) are available. We call them \emph{the nonlinear degrees of freedom}.

\end{abstract}


\section{Introduction}

Noncoherent communication systems in which neither the transmitter nor the receiver know the fading coefficients are of both theoretical and practical interest. The classical, intuitive approach towards these systems working in high SNR regime is training the receiver about the channel fading coefficients by transmitting some fixed symbols over a small period of time. The received signals at the training phase help the receiver gain some information about the channel fading coefficients. Knowing the fading coefficients, the receiver can recover the message from the received signal at the transmission phase.

The model of wireless channel of interest here is a time-varying, correlatively-changing fading channel. This is the model introduced and discussed in \cite{Liang}.

In time-varying MIMO fading channels with $n_t$ transmitting antennas and $n_r$ receiving antennas, the relationship between the transmitted signal at time $t$, $\mathbf{x}\tm{t}\in\mathcal{C}^{n_{t}}$ (the $t$-th column of $\mathbf{X} \in \mathcal{C}^{n_{t}\times T}$) and the noise-less received signal, $\mathbf{y}\tm{t}\in \mathcal{C}^{n_{r}}$ (the $t$-th column of $\mathbf{Y} \in \mathcal{C}^{n_{r}\times T}$) is characterized by the following equation

\begin{equation}
\label{eq:channel}
	\mathbf{y}\tm{t} = \mathbf{H}\tm{t} \mathbf{x}\tm{t}
\end{equation}

The random matrix $\mathbf{H}\tm{t} =[h_{m,n}\tm{t}] \in \mathcal{C}^{n_{r}\times n_{t}}$ contains the fading coefficients at time $t$.
The fading coefficient between $m$th transmit antenna and $n$th receive antenna, $h_{m,n}\tm{t}$, has normal complex gaussian distribution for $m=1,\cdots, n_{t}$ and $n=1,\cdots,n_{r}$. The fading coefficients between different pairs of transmitters and receivers are independent of each other.


The noisy received signal at the receiver is $\Y_{\text{noisy}}=\Y+\W$ where $\W \in \mathcal{C}^{n_r \times T}$ is the random IID complex gaussian noise at the receiver.

In the correlatively changing channel, fading coefficients change correlatively over time. The correlation matrix of fading coefficients between a pair of transmitter and receiver in a block of length $T$, denoted by $K_{\mathbf{H}}$, is of rank $Q$. 
Equivalently, all the fading coefficients between a pair of transmitter and receiver in a block of length $T$ are linear combinations of $Q$ statistically independent elements. $Q < T$ is the rank of the correlation matrix. The case $Q=1$ corresponds to the block fading model in which the fading coefficients do not change in the block of length $T$.

Define $\underline{\mathbf{h}}_{m,n}=[h_{m,n}\tm{1},\cdots,h_{m,n}\tm{T}]$  and $K_{\mathbf{H}}=\mathbb{E}[\underline{\mathbf{h}}^{\dagger}_{m,n} \underline{\mathbf{h}}_{m,n}] = \A^{\dagger} \A \in \mathcal{C}^{T \times T}$.
The vector  $\underline{\mathbf{s}}_{m,n}=[s^1_{m,n},\cdots, s^Q_{m,n}]$ contains the $Q$ statistically independent elements whose linear combinations give the elements of vector $\underline{\mathbf{h}}_{m,n}$.
The matrix $\A\in \mathcal{C}^{Q \times T}$, known at both Tx and Rx, gives the linear equations which specify the fading coefficients $h_{m,n}\tm{t}$ from the independent, gaussian distributed numbers $\underline{\mathbf{s}}_{m,n}$ as follows:

\begin{equation}
	\label{eq:whitening} h_{m,n}\tm{t}=\sum_{q=1}^Q A^q\tm{t} s^q_{m,n}
\end{equation}

Defining $A^q\tm{t}$ to be the element in $q$-th row and $t$-th column of matrix $\A$.


In the high SNR regime of fading channels, the measure of quality of interest is the degrees of freedom of the system. The DOF is defined as the pre-log factor in the first order approximation of capacity of the system in the high SNR regime. 

\[\mbox{DOF}=\lim_{\mbox{SNR}\to \infty}\frac{C(\mbox{SNR})}{\log \mbox{SNR}}\]

The DOF is interpreted as the number of dimensions in which communications can take place as SNR is increasing. 


The DOF of the system can be visualized in the following dimension counting argument:

Start with generating particular initial realization of the independent parameters of the channel ($\underline{\mathbf{s}}_{0,m,n}$) and the transmitted signals ($\X_0$). The noiseless received signal ($\Y_{0}$) would be derived from equations \eqref{eq:channel} and \eqref{eq:whitening}. 
Note that in instantiating the channel coefficients, only the independent parameters $\underline{\mathbf{s}}_{0,m,n}$ are realized and the fading coefficients are derived using equation \eqref{eq:whitening}.

Altering the realization of $\mathbf{H}\tm{t}$ and  $\X$ from $\mathbf{H}_0\tm{t}$ and $\X_0$ locally, we can move $\Y$ in a neighborhood around $\Y_{0}$. This neighborhood around $\Y_{0}$ is a subset of the $n_r T$ dimensional space reachable in the noisy version of received signal $\Y_{\text{noisy}}\in \mathcal{C}^{n_r\times T}$.

To find the number of DOF, we categorize the dimensions of the neighborhood reachable in \emph{noiseless received signal} into three categories:

\begin{enumerate}
	\item The dimensions of the neighborhood around $\Y_0$ that is reachable only by altering the realization of $\mathbf{H}\tm{t}$ from $\mathbf{H}_0\tm{t}$. Having fixed $\mathbf{H}_0\tm{t}$ and changing $\X$ from $\X_0$, $\Y$ can not change from $\Y_{0}$ along these dimensions.
	\item The dimensions of the neighborhood around $\Y_0$ that is reachable only by altering transmit signal $\X$ from $\X_0$.
	\item The dimensions of the neighborhood around $\Y_0$ that is reachable by altering transmit signal $\X$ from $\X_0$ or changing $\mathbf{H}\tm{t}$ from $\mathbf{H}_0\tm{t}$. Fixing either of transmitted signal or channel realization and changing the other one, $\Y$ can move from $\Y_0$ along these dimensions.
\end{enumerate}

Section~\ref{sec:DOF} gives a theorem which proves that in the high SNR regime, communications can take place only along the dimensions in the second category. It studies the mapping from the input signal space to the signal space of the noise-less received signal. It the states that if for some $X_0$ in the input signal space of the channel, there a $D$ dimensional neighborhood which is mapped to a $D$ dimensional neighborhood in the output signal space, and this mapping is one-to-one with probability one in this neighborhood, then the DOF of $D$ is achievable in this system.

Applying this argument to the flat fading channel  with $n_t \leq n_r$ in which $\Y=\mathbf{H}\X$, we are interested in the dimensions of the space where the signal $\Y$ can move by altering the realizations of $\mathbf{H}$ and $\X$.  Using the notation introduced in \cite{Zheng}, the subspace spanned by the rows of matrix $\Y$, $\Omega_{\Y}$, is the same as linear subspace spanned by the rows of matrix $\X$, $\Omega_{\X}$. 

Changing the realization of $\mathbf{H}$ does not change $\Omega_{\Y}$. Meaning that  $\Omega_{\Y}$ specifies the dimensions in $\Y$ which are \emph{only} reachable by the transmitted signal. Thus, this subspace of dimension $n_t(T-n_t)$ falls into the second category.
  The representation of the rows of matrix $\Y$ in the canonical basis of $\Omega_{\Y}$, ($\mathbf{C}_{\Y}\in \mathcal{C}^{n_r \times n_t}$) depends on both $\mathbf{C}_{\X}\in \mathcal{C}^{n_t \times n_t}$ and $\mathbf{H}\in \mathcal{C}^{n_r \times n_t}$. Thus, $n_t \times n_t$ dimensions in $\mathbf{C}_{\Y}$ are reachable by both $\mathbf{H}$ and $\X$ which fall into the third category. And $n_t(n_r-n_t)$ dimensions in $\mathbf{C}_{\Y}$ are only reachable by the different realizations of $\mathbf{H}$ which fall into the first category.

Trying to apply the same dimension counting argument to the received signals in correlatively changing channel, we analyze the noiseless received signal from $n$-th antenna. $\mathbf{y}_n \in \mathcal{C}^{1\times T}$ is the $n$-th row of matrix $\Y$.

\[\mathbf{y}_n=\sum_{m=1}^{n_t}\sum_{q=1}^{Q}s^q_{m,n} \underline{\A}^q \diag(\mathbf{x}_m)\]

Where $\underline{\A}^q$ is the $q$th row of matrix $\A$ and $\diag(\mathbf{x}_m)$ is a $T\times T$ matrix whose diagonal elements are the transmitted signals from $m$th antenna in a block of time.

We observe that the noiseless received signals from each antenna in a block of time, $\mathbf{y}_n$'s, live in a subspace $\F_{\A}(\X)=\spn\{\underline{\A}^q \diag(\mathbf{x}_m), \text{ for } m=1,\cdots,n_{t} \text{ and } q=1,\cdots,Q\}$. The nonlinear transform $\F_{\A}(\X)$ is a mapping from $\Omega_{\X}$ to a higher $n_t Q$ dimensional subspace, parameterized by the matrix $\A$. In the regime of interest where $n_r \geq n_t Q$, with probability one $\Omega_{\Y}=\F_{\A}(\X)$. The subset of the output space which are only reachable by the altering transmitted signal are a subset of $\F_{\A}{(\X)}$. But not all $n_tQ(T-n_t Q)$ dimensions of $\F_{\A}(\X)$ are reachable by changing $\X$. Studying this mapping over the manifolds of different dimensions, a mathematical tool is proposed which aims to count the number of dimensions in $\F_{\A}(\X)$ which are reachable only by altering $\X$, i.e., the DOF of the system.

In this paper, decoding algorithms for SIMO and MIMO systems in the regime when $n_tQ \leq \min(T-1,n_r)$ are proposed. These algorithms achieve the $n_t (1-n_t/T)$ DOF per symbol which is strictly larger than the one given in the conjecture in \cite{Liang}. Due to the nonlinearity of the mapping and decoding algorithms, we call the dimensions of the output subspace achieved by this method, \emph{nonlinear degrees of freedom}.

Having the classical training approach in mind, one might try to estimate the unknown parameters of the channel in each block and then communicate the message knowing the fading coefficients. This is the approach taken in \cite{Liang} where it is proved that this is the optimal strategy in terms of the achievable DOF in single-input single-output (SISO) systems. In a block of length $T$, $Q$ symbols are assigned to gather information about the fading coefficients in the training phase and $T-Q$ symbols are used to convey the message in the transmission phase. Thus there are  $(1-Q/T)$ DOF per symbol.

In the same paper, there is a conjecture about the MIMO systems which states that if $n_{t}<\min\{{n_{r}},T/2\}$, the pre-log factor of the system is $n_{t}(1-n_{t}Q/T)$. In a block of length $T$, there are $n_{t}^2 Q$ independent unknown elements which describe the fading coefficients in this block. Thus, $n_{t}^2Q$ symbols are assigned to gather information about the fading coefficients and $n_{t} T-n_{t}^2Q$ symbols are used to transmit information. The loss in the number of DOF due to the training is $n_t^2 Q$ which is proportional to the rank of the correlation matrix in this case. This conjecture is proved to be wrong for SIMO systems in \cite{Riegler} and \cite{Riegler2}. Our paper proves that this is not true for MIMO systems either and strictly higher number of DOF can be achieved using nonlinear decoding algorithms.

In \cite{Riegler}, SIMO systems are studied. Hironaka's theorem on resolution of singularities in algebraic geometry is used to prove that the pre-log factor of $(1-1/T)$ is achievable as long as $T>2Q-1$ under some constraints over the correlation matrix of the fading coefficients. We prove the achievability of $(1-1/T)$ DOF when $Q \leq \min(T-1,n_r)$ in SIMO systems under some mild conditions and give the proper nonlinear decoding algorithm. The constraints under which these DOF are achievable are much milder than the ones given in \cite{Riegler}. 

Also the achievability of $1-\lceil\frac{Q}{n_r}\rceil/T$ DOF per symbol for the general number of received antennas is given in this paper.

For MIMO systems the decoding algorithm and some mild sufficient conditions to achieve $n_t(1-n_t/T)$ DOF per transmitted symbol in the regime when $n_tQ\leq \min(T-n_t,n_r)$ is given in section \ref{sec:MIMO}.

\section{DOF as the Dimensionality }

\label{sec:DOF}

\begin{theorem}
	If for some $\X_0$ in the input space of the communication channel, there is a $D$ dimensional  neighborhood in input space which is mapped to a $D$ dimensional neighborhood in the noise-less output space and this mapping is one-to-one with probability one in this space, then the degrees of freedom $D$ is achievable in this system. 
\end{theorem}

To achieve the $D$ degrees of freedom in this neighborhood, QAM modulation is performed in each of the $D$ dimensions of input space which is conserved in output space. Define $d_{\text{min,x}}$ to be the  minimum distance of the codewords in input space. In the communication channel with signal to noise ratio $SNR$, we can assume the input power constraint implies $\mathbb{E}[\| x\|^2]\leq 1$ and stationary noise has power spectral density $1/\text{SNR}$. 

Define $d_{\text{min,y}}$ as the minimum distance between noiseless received codewords in the output space. Since the mapping is one-to-one with probability one in this space, the eigenvalues of the Jacobian of this mapping is strictly positive with probability one. So with probability $1-\epsilon$, the minimum eigenvalue of the Jacobian of the mapping is greater than $\sigma_0$. Thus, with probability $1-\epsilon$ we would have $d_{\text{min,y}}\geq d_{\text{min,x}} \sigma_0$. 

In the fading channel with the Rayleigh fading coefficients and noise power density $1/\text{SNR}$, the probability of error vanishes as long as $d_{\text{min,y}} \gg 1/{\sqrt{\text{SNR}}}$. Thus, the probability of error vanishes as long as $d_{\text{min,x}} \sigma_0 \gg 1/{\sqrt{\text{SNR}}}$. Setting $d_{\text{min,x}}=\frac{1}{\sigma_0 \text{SNR}^{1/2-\delta}}$, the probability of error vanishes. The power constraint implies that in each dimension, QAM would give $ (2/d_{\text{min,x}})^2$ codewords. Thus, with probability $1-\epsilon $ the total number of codewords would be $(2/d_{\text{min,x}})^{2D}=(2 \sigma_0 \text{SNR}^{1/2-\delta})^{2D}$ and the achievable rate is $(1-\epsilon)(1-2\delta)D\log(SNR)+o(\log SNR)$.

\section{A Simple Example}

Assume a SIMO system with parameters $T=3, Q=2, n_r=2$ and ${n_{t}}=1$ in which the fading coefficients at $t=1,2$ are statistically independent of each other and the coefficient at $t=3$ is a linear combination of the coefficients at $t=1$ and $2$.
Thus, $\mathbf{h}_1=\left[ h_1\tm{1} , h_1\tm{2}, h_1\tm{3} \right]$ and $\mathbf{h}_2=\left[ h_2\tm{1} , h_2\tm{2}, h_2\tm{3} \right]$ where $h_1\tm{3} = \alpha h_1\tm{1}+ \beta h_1\tm{2}$ and $h_2\tm{3} = \alpha h_2\tm{1}+ \beta h_2\tm{2}$ for some $\alpha, \beta \in \mathcal{C}$. The correlation matrix has rank $2$ in this case 

\[K_H=\begin{bmatrix}  1 & 0 \\ 0 & 1 \\ \alpha^* & \beta^* \end{bmatrix} \begin{bmatrix}  1 & 0 & \alpha \\ 0 & 1 & \beta  \end{bmatrix}\].

The transmitted signal in a block of time is $\mathbf{x}=\left[x\tm{1} , x\tm{2}, x\tm{3}\right]$. The noise-free received signal in the high SNR regime is 

\begin{eqnarray*} \mathbf{y}_1 & = & \left[ h_1\tm{1} x\tm{1} , h_1\tm{2} x\tm{2}, (\alpha h_1\tm{1}+ \beta h_1\tm{2})x\tm{3} \right] \\
\mathbf{y}_2 & = & \left[ h_2\tm{1} x\tm{1} , h_2\tm{2} x\tm{2}, (\alpha h_2\tm{1}+ \beta h_2\tm{2})x\tm{3} \right]
\end{eqnarray*}

It is obvious that having $\mathbf{y}_1$ and $\mathbf{y}_2$, the transmitted signals can not be recovered uniquely without any side information about the realization of the channel. 

The classical approach towards the non-coherent communications would transmit training signals over $t=1,2$ to estimate the unknown fading coefficients ($\{h_1\tm{1} , h_1\tm{2}, h_2\tm{1} , h_2\tm{2}\}$) and the message would be transmitted at time $t=3$. Thus the degrees of freedom of the channel would be $1/3$.

But looking more closely, we realize that if training is performed only at $t=1$, we can recover the transmitted messages at $t=2,3$ without first estimating all the fading coefficients explicitly. Assume that $x\tm{1}=1$.

At the receiver, both antennas perform a nonlinear calculation and divide their received signals by the $y_1\tm{1}$ and $y_2\tm{1}$, 

\begin{eqnarray*} \mathbf{y}_1/y_1\tm{1} =  \left[ 1  , h_1\tm{2}/h_1\tm{1} x\tm{2} , (\alpha + \beta h_1\tm{2}/h_1\tm{1}) x\tm{3} \right] \\
	 \mathbf{y}_2/y_2\tm{1} = \left[ 1  , h_2\tm{2}/h_2\tm{1} x\tm{2} , (\alpha + \beta h_2\tm{2}/h_2\tm{1}) x\tm{3} \right]
\end{eqnarray*}

Now the transmitted signals at $t=2,3$ can be decoded by solving the $4$ non-linear equations with $4$ unknowns (i.e., $x\tm{2},x\tm{3},h_1\tm{2}/h_1\tm{2}, h_2\tm{2}/h_2\tm{1}$). We observe that even though after finding the transmitted signals, we can estimate all the fading coefficients, but we didn't need to explicitly estimate them before decoding the transmitted signals. The above nonlinear approach enabled us to achieve $(1-1/T)=2/3$ degrees of freedom per symbol which is strictly larger than the classical training approach which tries to estimate all the fading coefficients explicitly before the communication phase.



\section{SIMO Systems}


\label{sec:SIMO}
In the model of SIMO systems of interest, there is one transmit antenna and ${n_{r}}$ receive antennas. $x\tm{t}$ for $t=1,\cdots,T$ is the transmitted signal at time $t$. $y_n\tm{t}$ for $n=1,\cdots {n_{r}}$ denotes the noise-free signal at $n{\text{th}}$ receive antenna at time $t$. Also, $h_n\tm{t}$ is the fading coefficient between the transmitter and the $n$th receive antenna at time $t$. We define $\mathbf{x}=[x\tm{1},x\tm{2},\cdots,x\tm{T}]$, $\mathbf{y}_n=[y_n\tm{1},y_n\tm{2},\cdots,y_n\tm{T}]$ and $\underline{\mathbf{h}}_n=[h_n\tm{1},h_n\tm{2},\cdots,h_n\tm{T}]$. 

The correlation matrix $K_H=\A^{\dagger} \A$, known at the receivers as channel side information has rank $Q<T$. Thus, $h_n\tm{t}$, can be written as $h_n\tm{t}=\sum_{q=1}^{Q}A^q\tm{t}s^q_n$ where $s^q_n$'s are IID normal Gaussian distributed.






The noise-free received signals from the channel can be expressed as 



\[\mathbf{y}_n=\sum_{q=1}^Q s^q_n \left[\underline{\A}^q \diag(\mathbf{x})\right]\]

The noise-less received signal at each antenna, $\mathbf{y}_n$, is living in a $Q$ dimensional linear subspace $\F_{\A}(\mathbf{x})=\spn\{\underline{\A}^q \diag(\mathbf{x}) \text{ for }q=1,\cdots,Q\}$.

$\F_{\A}(\mathbf{x})$ is a nonlinear mapping from vector $\mathbf{x}$ to a $Q$ dimensional subspace in $\mathcal{C}^T$. Since $\mathbf{y}_n \in \F_{\A}(\mathbf{x})$ for all $n=1,\cdots,{n_{r}}$, looking at the received signals in all the antennas in the high SNR regime, $\F_{\A}(\mathbf{x})$ can be recovered with probability one as long as ${n_{r}}\geq Q$. In this regime, we have $\F_{\A}(\mathbf{x}) = \spn\{\mathbf{y}_1,\mathbf{y}_2,\cdots,\mathbf{y}_{n_{r}}\}$.

We want to know which dimensions in $\F_{\A}(\mathbf{x})$ are reachable by changing the transmitted signal $\mathbf{x}$.

Firstly, it is observed that for any $\alpha \in \mathcal{C}$, $\F_{\A}(\alpha \mathbf{x})=\F_{\A}(\mathbf{x})$. Thus, we need to lose at least one degree of freedom in the representation of the vector $\mathbf{x}$ to be able to recover it from $\F_{\A}(\mathbf{x})$ uniquely. Having this constraint, $\F_{\A}(\mathbf{x})$ is a mapping over the manifolds from a one-dimensional linear subspace in $\mathcal{C}^T$ to a $Q$ dimensional linear subspace.
The loss of one degree of freedom can be in form of normalization of the transmitted power or equivalently training (e.g., $x\tm{T}=1$). Proving that by training one degree of freedom, the transmitted message is recoverable with probability one, we know that there are $T-1$ degrees of freedom in a block of length $T$ in this regime.

In the high SNR regime, $\F_{\A}(\mathbf{x})$ can be restored from the noise-free received signals with probability one as long as $Q \leq n_r$. In the receiver, after building the canonical form of $\F_{\A}(\mathbf{x})$, the decoding algorithm tries to recover the transmitted signal $\mathbf{x}$. We can prove that the canonical form is a bijective function of the transmitted signal with probability one. This being true, a decoding algorithm is proposed to recover the transmitted signal.

\subsection{Change of Coordinates} 
\label{sec:coordinates}
Following the notation in \cite{Zheng}, each linear subspace of dimension $L$ in $\mathcal{C}^T$ can be represented as span of $L$ linearly independent vectors in the rows of a matrix  $\R \in \mathcal{C}^{L\times T}$. The same subspace is represented by choosing any non-singular matrix $\mathbf{C}\in \mathcal{C}^{L\times L}$ and constructing matrix $\B \in \mathcal{C}^{L\times T}$ such that $\R=\mathbf{C}\B$. The matrix $\mathbf{C}_{\R}$ can be chosen such that $\B\tm{1:L}=\I_{L}$, where $\B\tm{1:L}\in\mathcal{C}^{L \times L}$ is the submatrix of first $L$ columns of matrix $\B$ and $\I_{L}$ is the identity matrix of size $L$.



We call matrix $\B$ \emph{the canonical representation} of this linear subspace of dimension $L$ in $\mathcal{C}^T$. Choosing $\mathbf{C}_{\R}=\R\tm{1:L}$ as the first $L$ columns of matrix $\R$, we construct $\B= \C^{-1}_{\R} \R$.

\subsection{Mapping over the Manifolds}

As mentioned,  $\F_{\A}(\mathbf{x})$ is a mapping over the manifolds. In order to study this mapping, we use the canonical form of the linear subspaces. By using the canonical form for both the input and output of the transform,  $\F_{\A}(\mathbf{x})$ can equivalently be represented as $\F_{\A}:\mathcal{C}^{T-1}\to \mathcal{C}^{Q(T-Q)}$. 
 
For simplicity of decoding algorithm, the training is performed as constraining the transmitted signal $\mathbf{x}$ so that $x\tm{T}=1$. So the input of the transform $\F_{\A}$ is $[x\tm{1},x\tm{2},\cdots,x\tm{T-1}]\in \mathcal{C}^{T-1}$. The output of the transform is the canonical form of $\F_{\A}(\mathbf{x})$ as introduced in section \ref{sec:coordinates}. Thus the parameters, determining the output of the transform is the non-trivial components of matrix $\B$ which will be described below. We will show that we can choose a set of dimension $T-1$ of the parameters defining the output and form a bijective transform between the input and output with probability one.

We know that $\F_{\A}(\mathbf{x})$ can be represented as the span of rows of matrix $\R=\A \diag{(\mathbf{x})}$. In order to form the canonical representation of $\F_{\A}(\mathbf{x})$, we choose $\mathbf{C}_{\R}$ as the following,

\[\mathbf{C}_{\R} =\R\tm{1:Q} =  \A\tm{1:Q} \diag(\mathbf{x}\tm{1:Q})\]

Where $\mathbf{x}\tm{1:Q}\in \mathcal{C}^{Q}$ is vector of first $Q$ elements of vector $\mathbf{x}$. We define $\underline{\A}\tm{t}$ as the $t$-th column of matrix $\A$. Defining $\underline{\R}\tm{t}$ and $\underline{\B}\tm{t}$ similarly, $\underline{\R}\tm{t}=\underline{\A}\tm{t} x\tm{t}$. Since $\B={\mathbf{C}^{-1}_{\R}} \R$, we know that for all $t=Q+1,\cdots,T$:

\begin{eqnarray*}
	\label{eq:SIMO_Bt1}
	\underline{\B}\tm{t}
	& = &  \left( \A\tm{1:Q} \diag(\mathbf{x}\tm{1:Q}) \right)^{-1}  \underline{\A}\tm{t} x\tm{t}\\
	& = &  x\tm{t} \left(\diag(\mathbf{x}\tm{1:Q}) \right)^{-1} \left( \A\tm{1:Q} \right)^{-1} \underline{\A}\tm{t} \end{eqnarray*}

Defining the vector $\underline{\mathbf{E}}\tm{t}= \left( \A\tm{1:Q} \right)^{-1} \underline{\A}\tm{t}$, known as the channel side information at the receiver for any $t=Q+1,\cdots,T$ and $q=1,\cdots,Q$, 

\begin{equation}
	\label{eq:SIMO-decode}B^q\tm{t}=E^q\tm{t} x\tm{t}/x\tm{q}
\end{equation}

\subsection{Decoding Algorithm}
\label{sec:decoding_SIMO}

The training $x\tm{T}=1$ is assumed for the transmitted signal. We also assume that $Q<T$ and $Q \leq n_r$.

In the high SNR regime, the effect of noise is neglected. Thus the decoding algorithm performs as follows:

\begin{enumerate}
	\item Construct $\hat{\F}_{\A}(\mathbf{x})$ as the span of rows the low rank approximation of $\Y_{\text{noisy}}$. It could be formed to be the span of the $Q$ right singular vectors of $\Y_{\text{noisy}}$ corresponding to its $Q$ largest singular values.
	\item Construct matrix $\hat{\B}$ as the canonical form of $\hat{\F_{\A}}(\textbf{x})$ such that $B\tm{1:Q}=\I_{Q}$.
	\item Compute $\underline{E}\tm{t}=\left( \A\tm{1:Q} \right)^{-1} \underline{\A}\tm{t}$ for $t>Q$.
	\item For $q=1,\cdots,Q$, use \eqref{eq:SIMO-decode} as $\hat{x}\tm{q}=E^q\tm{T}/\hat{B}^q\tm{T}$ 
		\item  Having $\hat{x}\tm{1}$ for $Q+1 \leq t < T$ , use \eqref{eq:SIMO-decode} as
		\[\hat{x}\tm{t}=\hat{x}\tm{1}  \frac{\hat{B}^1\tm{t}}{E^1\tm{t}}\]
\end{enumerate}

\subsection{Recovery Conditions}
\label{sec:constraint_SIMO}
Looking at the decoding algorithm with the above assumptions, the Jacobian of the mapping between the manifolds can be computed. The following theorem is the result of this computation.

\begin{theorem}
	The signal recovery at the receiver succeeds if the following conditions are satisfied:
\begin{itemize}
	\item $x\tm{q}\neq 0$ for any $1 \leq q \leq Q$. This situation can be avoided with probability one by simply assuming any continuous distribution over the transmitted signals.
	\item $E^q\tm{T}\neq0$ for any $1 \leq q\leq Q$.
	\item $E^1\tm{t}\neq0$ for any $Q+1 \leq t\leq T-1$.
\end{itemize}
\end{theorem}
Having defined $\underline{\mathbf{E}}\tm{t}= \left( \A\tm{1:Q} \right)^{-1} \underline{\A}\tm{t}$, $E^q\tm{t}\neq0$ if and only if

\[\left|\begin{matrix} 
A^1\tm{1} & \cdots & A^1\tm{q-1} &  A^1\tm{t} &  A^1\tm{q+1} & \dots & A^1\tm{Q}\\ 
A^2\tm{1} & \cdots & A^2\tm{q-1} &  A^2\tm{t} &   A^2\tm{q+1} & \dots & A^2\tm{Q} \\
\vdots & &\vdots& \vdots & \vdots & & \vdots\\
A^Q\tm{1} & \cdots  & A^Q\tm{q-1} &  A^Q\tm{t} & A^Q\tm{q+1}  & \dots & A^Q\tm{Q}\\
\end{matrix}\right|\neq0\]

Therefore, to be able to perform the decoding process successfully, in addition to having a continuous distribution over the transmitted signals, we should guarantee that the choice of $Q$ columns of the matrix $\A$, corresponding to cases 2 and 3 are linearly independent of each other.

This recovery condition is less restrictive than the one mentioned in \cite{Riegler} which requires that any choice of $Q$ columns of matrix $\A$ are linearly independent of each other. In \cite{Riegler}, resolution of singularities is used to prove that the mapping introduced is one-to-one almost everywhere. It is proved that the expectation of logarithm of determinant of Jacobian is not $-\infty$. 

Our work is based on introducing a mapping from a lower dimensional manifold to higher dimensional one. A mathematical tool is introduced which studies of this mapping using the canonical representation of the manifolds. This study can compute the actual value of the Jacobian of the transform and proves the achievability of nonlinear degrees of freedom with probability one under the specified constraints.

\subsection{Systems with $n_r<Q$ }
In SIMO systems with $n_r<Q$, the mentioned decoding algorithm is not going to be feasible due to the fact that the subspace $\F_{\A}(\mathbf{x})$ can not be estimated from the received signals in $n_r$ receive antennas. We define the matrix $\G\in \mathcal{C}^{n_r\times T} $ to be the canonical form of the linear subspace spanned by the received signals from $n_r$ antennas. We know that each row of $G$, $G^n\in\mathcal{C}^{1\times T}$ is in $\F_{\A}(\mathbf{x})$, i.e., $G^n\in \F_{\A}(\mathbf{x})$. Since the matrix $B$ is the canonical representation of $\F_{\A}(\mathbf{x})$ where $\B\tm{1:Q}=\I_{Q}$, we would know that

\[\G=\G\tm{1:Q}\B\]

Note that the matrix $\G$ is estimated at the receiver as the canonical representation of $n_r$ dimensional linear subspace spanned by the noise-less received signals. Thus, for $t=Q+1,...,T$, we would know that

\begin{eqnarray*}
\G\tm{t} & = & \G\tm{1:Q}\B\tm{t}\\
& = & \G\tm{1:Q} \left(\diag(\mathbf{x}\tm{1:Q}) \right)^{-1} \underline{\mathbf{E}}\tm{t} x\tm{t}\\
& = & \G\tm{1:Q} \diag(\underline{\mathbf{E}}\tm{t}) [x\tm{t}/x\tm{1}, \cdots, x\tm{t}/x\tm{Q}]^T
\end{eqnarray*}

If $n_r=Q$ and assuming $x\tm{T}=1$, looking at the above equation for $t=T$, $Q$ linear equations in $1/x\tm{q}$ for $q=1,\cdots,Q$ construct the nonlinear phase of the decoding algorithm. But in the case $n_r<Q$, this is an underspecified set of equations. Thus, more training should be done to be able to solve this system of equations. Assuming $x\tm{T}=x\tm{T-1}=\cdots=x\tm{T-\lceil\frac{Q}{n_r}\rceil+1}=1$, this (probably over-specified) system of equations with $n_r\lceil\frac{Q}{n_r}\rceil$ equations and $Q$ variables can be solved to give $1/x\tm{q}$ for $q=1,\cdots,Q$.

\[\begin{bmatrix}\G\tm{T}\\ \vdots \\ \G\tm{T-\lceil\frac{Q}{n_r}\rceil+1} \end{bmatrix}=
\begin{bmatrix} \G\tm{1:Q} \diag(\underline{\mathbf{E}}\tm{T}) \\ \vdots \\ \G\tm{1:Q} \diag(\underline{\mathbf{E}}\tm{T-\lceil\frac{Q}{n_r}\rceil+1})\end{bmatrix} \begin{bmatrix} 1/x\tm{1} \\ \vdots \\ 1/x\tm{Q}\end{bmatrix}\]

The number of degrees of freedom in correlatively changing SIMO systems is $1-\lceil\frac{Q}{n_r}\rceil/T$.

\section{MIMO Systems}
\label{sec:MIMO}
Similar to SIMO systems, the channel in time varying fading systems can be modeled as nonlinear mapping from the transmitted signals to the noise-less received signals. In correlatively changing channels, This nonlinear function is a mapping over the manifolds from a lower dimensional linear subspace to higher dimensional subspace. In order to find the number of degrees of freedom of the system, the high SNR received signals are used to construct the output linear subspace of this mapping. Then, the canonical representation of this subspace is constructed. The goal is to find as many number of dimensions of the input signal as possible from this canonical representation.

As shown previously the noiseless received signal from $n$-th antenna in a block of time of length $T$ is

\[\mathbf{y}_n=\sum_{q=1}^{Q}\sum_{m=1}^{n_{t}} s^q_{m,n} \left[\underline{\A}^q  \diag{(\mathbf{x}_m})\right]\]

Thus, the linear subspace carrying the message at the receiver is 

\begin{eqnarray*}\F_{\A}(\X) = \spn\{ \underline{\A}^q \diag(\mathbf{x}_m) &\text{ for } q=1,\dots,Q \\ &\text{ and } m=1,\cdots,{n_{t}}\}
\end{eqnarray*}

For all $n=1,\cdots,n_r$, $\mathbf{y}_n\in \F_{\A}(\X)$.

It is easily seen that $\F_{\A}(\X)$ is a mapping from $\Omega_{\X}$ to an $n_t Q$ dimensional linear subspace. Thus, the message is transmitted through the linear subspace spanning the rows of matrix $\X$. By fixing any $n_t$ columns of matrix $\X$, this linear subspace is specified uniquely. This can be done by transmitting the training signals in any $n_t$  chosen time slots in the block of time of length $T$.

The matrix $\R \in \mathcal{C}^{{n_{t}}Q \times T}$ whose rows span $\F_{\A}(\X)$ is 

\begin{equation}
	\label{eq:R-MIMO}
	\R=\begin{bmatrix}
\A \diag(\mathbf{x}_1)\\
\A \diag(\mathbf{x}_2)\\
\vdots\\
\A \diag(\mathbf{x}_{n_{t}})\\
\end{bmatrix}
\end{equation}

Defining $\A\tm{1:n_{t}Q}\in \mathcal{C}^{Q\times n_t Q}$ and  ${\mathbf{x}}_m\tm{1:n_t Q}$ similar to the SIMO case, the first ${n_{t}}Q$ columns of matrix $\R$ would be

\[\mathbf{C}_{\R}= \R\tm{1:n_tQ} = \begin{bmatrix}
\A\tm{1:{n_{t}}Q} \diag(\mathbf{x}_1\tm{1:{n_{t}}Q})\\
\A\tm{1:{n_{t}}Q} \diag(\mathbf{x}_2\tm{1:{n_{t}}Q})\\
\vdots\\
\A\tm{1:{n_{t}}Q} \diag(\mathbf{x}_{n_{t}}\tm{1:{n_{t}}Q})\\
\end{bmatrix}\]

Since $\B={\mathbf{C}_{\R}}^{-1}\R$ we know that $\underline{\R}\tm{t}={\mathbf{C}_{\R}} \underline{\B}\tm{t}$ for all $t={n_{t}}Q+1,\cdots,T$.

Thus looking at the $t$-th column of matrix $\R$ as defined in~\eqref{eq:R-MIMO}, for $m=1,\cdots,{n_{t}}$ and $t > n_tQ$, we would know

\begin{eqnarray}\underline{\A}\tm{t} x_m\tm{t}  & = & \A\tm{1:{n_{t}}Q}\diag( \mathbf{x}_m\tm{1:n_t Q})\underline{\B}\tm{t} \notag \\
	 & = & \A\tm{1:{n_{t}}Q}\diag(\underline{\B}\tm{t}) \mathbf{x}^T_m\tm{1:n_t Q} \label{eq:MIMO1}
\end{eqnarray}

Since with the above choice of $\mathbf{C}_{\R}$, $\B\tm{1:n_t Q}=I_{n_t Q}$, this relationship between $x_m\tm{t}$ and $\underline{\B}\tm{t}$ is trivial for $t\leq n_t Q$. 

For $t> n_t Q$, we use a two phase decoding algorithm, which uses the training signals in the first nonlinear phase to get information about $\X\tm{1:n_t Q}$. In the linear phase of the decoding algorithm, $\hat{\X}\tm{1:n_t Q}$ is used to estimate the remaining transmitted signals.

The nonlinear phase of the algorithm is preformed for $t=n_t Q+1,\cdots,n_t(Q+1)$. We choose the training signal to have the form $\X\tm{n_tQ+1:n_t(Q+1)}=\I_{n_t}$. 

Writing the equation \ref{eq:MIMO1} for all $n_t Q<t \leq n_t(Q+1)$ in a matrix, for $m=1,\cdots,n_t$, we would have,

\begin{multline}
	\label{eq:MIMO-nonlinear}
\begin{bmatrix}
\underline{\A}\tm{n_t Q+1} x_m\tm{n_t Q+1}\\
\underline{\A}\tm{n_t Q+2} x_m\tm{n_t Q+2} \\
\vdots\\
\underline{\A}\tm{n_t(Q+1))} x_m\tm{n_t(Q+1))}\\
\end{bmatrix}=\\
 \begin{bmatrix}
\A\tm{1:{n_{t}}Q} \diag(\underline{\B}\tm{n_t Q+1})\\
\A\tm{1:{n_{t}}Q} \diag(\underline{\B}\tm{n_t Q+2})\\
\vdots\\
\A\tm{1:{n_{t}}Q} \diag(\underline{\B}\tm{n_t(Q+1)})\\
\end{bmatrix} \mathbf{x}^T_m\tm{1:{n_{t}}Q}
\end{multline}

The left hand side of the above equation is known at both the transmitter and the receiver as a result of the training. Having constructed the matrix $\B$, the inverse of the transform in equation \ref{eq:MIMO-nonlinear} can be used to estimate $\mathbf{\hat{x}}_m\tm{1:{n_{t}}Q}$. 

\subsection{Decoding Algorithm}

So the decoding algorithm at the receiver would perform as follows

\begin{enumerate}
	\item Construct $\hat{\F_{\A}}(\X)$ of dimension $n_t Q$ as the span of the rows of the low rank approximation of $\Omega_{\Y_{\text{noisy}}}$ similar to the SIMO case.
	\item Construct matrix $\hat{\B}$ as the canonical representation of $\hat{\F}_{\A}(\X)$.
	\item Use equation \eqref{eq:MIMO-nonlinear} to recover $\hat{\X}\tm{1:{n_{t}}Q}$.
	\item For $m=1,\cdots,{n_{t}}$ and  ${n_{t}}(Q+1) < t \leq T$, having $\hat{\X}\tm{1:{n_{t}}Q}$, to recover $\hat{x}_m\tm{t}$, use 
	\[ \hat{x}_m\tm{t} = \underline{\A}^1\tm{1:{n_{t}}Q}\diag(\hat{\underline{\B}}\tm{t}) \hat{\mathbf{x}}^T_m\tm{1:{n_{t}}Q}/A^1\tm{t}\] 
\end{enumerate}

\subsection{Recovery Conditions}

\begin{theorem}
	In correlatively changing fading MIMO systems in the regime when $n_t Q\leq \min(n_r,T-n_t)$ the number of DOF of $n_t(1-n_t/T)$ per transmitted symbol is achievable under the following sufficient conditions.

These conditions will provide the recovery of the transmitted message with probability one at the receiver using the above decoding algorithm.

\begin{itemize}
	\item The transmitted signal should have continuous distribution over linear subspaces of dimension $n_t$. 
	\item Any $Q$ columns of the matrix $\A\tm{1:n_t(Q+1)}$ should be linearly independent of each other.
\end{itemize}
\end{theorem}

The proof consists of two main parts. The first part proves that satisfying the above constraints, in the canonical representation of $\F_{\A}(\X)$, $B^q\tm{t} \neq 0$ with probability one for $n_tQ<t \leq n_t(Q+1)$.

In the second part, it is proved that in the nonlinear phase of the decoding algorithm, the above constraints are the sufficient conditions to have a locally surjective nonlinear transform from $\mathcal{C}^{n_tQ}$ to $\mathcal{C}^{n_t Q}$ which gives the $n_tQ$ nonlinear degrees of freedom that is achieved through the nonlinear decoding.

\section{Conclusion}
In this paper, using the dimension counting argument, the degrees of freedom of correlatively changing fading channel over SIMO and MIMO systems were analyzed when $n_t Q \leq\min(n_r,T-n_t)$. It is shown that in this channel, in the high SNR regime, information is transmitted through a nonlinear transform from the linear subspace spanned by the transmitted signals to the subspace spanned by the received signals. This transform is a mapping over the Grassman manifold of dimension $n_t$ in $\mathcal{C}^{T}$ to another manifold of dimension $n_t Q$ in $\mathcal{C}^T$. Analysis of the dimensions which are only reachable by the transmitted signal gives us $n_t(T-n_t)$ nonlinear degrees of freedom in a block of length $T$. This number is the same as the number of degrees of freedom in flat fading channel in the same regime. This shows that using proper nonlinear decoding techniques, correlatively changing channels can achieve the same pre-log factor as the flat fading channels.


\end{document}